\documentclass[11pt,a4paper]{article}
\pdfoutput=1
\usepackage{jheppub}
\usepackage{mathrsfs}
\usepackage{dsfont}

\title{Low-Lying Cosmic String Spectrum and Background Fields in Effective String Theory}

\author{Nicholas Agia}

\affiliation{Carnegie Mellon University, \\ 5000 Forbes Avenue, Pittsburgh, USA}

\emailAdd{nicholaa@andrew.cmu.edu}

\abstract{We present a detailed analysis of the cosmic string spectrum. Explicit solutions are numerically found using Mathematica and presented here for the lowest-lying supported modes. Most of the emphasis is on the Nambu-Goldstone modes and the least massive excitation, the latter of which is shown to be the scalar breather mode. We address the possibility of pseudoscalar excitations by adding suitable interactions to the string and show that it is possible to have a least massive pseudoscalar bound state with only bosonic fields. We finally show how certain interactions in the bulk UV theory give rise to background field interactions in the effective string theory.}


\begin{document} 
\maketitle
\flushbottom

\section{Introduction and Summary}
\hspace*{1em} The cosmic string as a topological soliton has been studied in different contexts quite extensively over the past several decades, a classic paper being Nielsen and Olesen's \cite{Nielsen and Olesen}. Vortices, even Abelian ones, are seemingly ubiquitous in physics, appearing widely from Abrikosov vortices in superconductors \cite{Abrikosov} to more exotic so-called Alice strings with nonlocalized Cheshire charge \cite{Alice strings}. Despite its history, however, we were not able to find an explicit presentation of the spectrum of particles supported by the Abelian string and its ramifications for its effective string theory. This paper aims to fill such a gap. In addition, a recent finding in the study of low energy QCD flux tube spectra provides strong evidence for the somewhat surprising result of the lowest-lying massive mode being given by a pseudoscalar resonance \cite{DFG1,DFG2}. This leads one to wonder to what extent this behavior appears in other theories. In addition, this light resonance in QCD, dubbed the \emph{worldsheet axion}, needs to be understood better both in QCD itself and in general.  As a step toward understanding these issues, it is instructive to study perhaps the simplest theory involving vortices, the Abelian Higgs model. 

In this paper we consider a vortex in a $(3+1)$-dimensional $U(1)$ gauge model, the field content being a complex scalar $\phi$ and a gauge field $A_{\mu}$. We focus on the dynamics of a long, straight cosmic string. The classical background scalar and gauge profiles for the string are well known; while their equations do not admit any known analytic solutions, it is a simple matter to numerically solve them \cite{Rubakov}. This leads to profiles shown, for example, in Figure \ref{background} in the following section. We then consider the fluctuations about this static background and calculate the low-lying spectrum of particles supported by the string in perturbation theory, since it seems this explicit calculation is not in the literature. As mentioned in the preceding paragraph, it is of interest to consider the discrete transformational properties of the least massive excitation, particularly those of parity. After explicitly calculating the exact form of the Nambu-Goldstone modes of the cosmic string, we show that the cosmic string's least massive excitation numerically determined corresponds to the scalar breather mode. Hence the spectrum of this Abelian theory begins with a scalar. 

It is straightforward to carry out this analysis in perturbation theory for successively more massive modes. We do this numerically and find nothing unexpected. A much more interesting story plays out when one couples external scalars and pseudoscalars to the cosmic string. We show that if we couple a pseudoscalar via a topological $F \wedge F$ term as well as a quartic coupling to the Higgs field, the least massive excitation is necessarily a pseudoscalar state over all ranges of the various couplings. This is a highly nontrivial result, especially since it depends critically on the presence of the $F \wedge F$ interaction. We find that if the $F \wedge F$ coupling is present, the least massive mode is the pseudoscalar no matter how small the coupling is; however, if the coupling vanishes exactly, it may be the scalar or the pseudoscalar.

Lastly, we consider the low-energy effective string theory. It turns out that the above topological coupling descends in this limit to a coherent Kalb-Ramond background, at least to tree-level. Matching to the effective theory at one-loop order then gives rise to a coupling of the pseudoscalar to the string self-intersection number. Following \cite{DFG1} and \cite{DFG2}, we call this coupling a \emph{worldsheet axion interaction}. 

The paper is organized as follows. In Section \ref{string equations}, we give the mode equations for fluctuations about the static cosmic string background. We also develop the gauge fixing and Fourier and spectral decompositions to be used to conveniently calculate the massive spectrum. In Section \ref{low-lying}, we present the solutions found for the lowest-lying excitations, namely the two massless and the single least-massive modes. Then in Section \ref{externals}, we introduce an external spinless field and couple it to the string. First, this is coupled only via a quartic interaction with the Higgs fields; then, the situation is analyzed by introducing the additional $F \wedge F$ interaction, pointing out the important difference in the string's mass spectrum. Finally, in Section \ref{effective string theory}, we take the UV theories considered in Section \ref{externals} and perform the matching to the effective string theory for the pseudoscalar interaction at tree-level and one-loop order. These illustrate how the long string can couple to a background Kalb-Ramond field and a worldsheet axion, respectively. Some additional details are included in an appendix useful to reproduce some of the calculations below. To be as self-consistent as possible, there is also a very brief account in the appendix of the appearance of the Abelian vortex in classical gauge theory and how the effective Nambu-Goto action appears from the coset construction of spontaneous symmetry breaking.

\section{Fluctuations about Cosmic String Background} \label{string equations}

\begin{figure}
\centering
\includegraphics[scale=0.65]{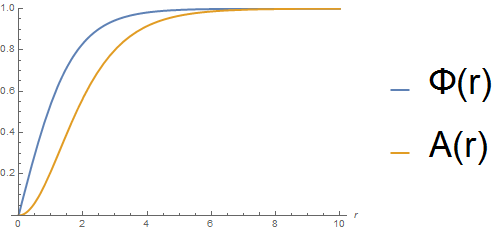}
\caption{The standard profiles for the cosmic string background, defined by the Ansatz \eqref{Ansatz}. As with all other numerical solutions presented in this paper, the explicit solutions were obtained using $m_H = m_V = 1$. These form factors also satisfy the BPS equations \eqref{BPS 1} and \eqref{BPS 2}.}
\label{background}
\end{figure}

The Abelian Higgs model is very familiar. For a review of the classical gauge theory, see for example \cite{Rubakov}. Our convention for the minimally coupled potential is
\begin{equation} V(|\phi|) = \frac{\lambda}{2}\left(\phi^* \phi - \frac{\mu^2}{\lambda}\right)^2. \end{equation}
We then take the soliton to be centered at the origin and extend infinitely in the $z$-direction via the background Ansatz
\begin{equation}\label{Ansatz} \phi_{cl} = \frac{\mu}{\sqrt{\lambda}}e^{i\varphi}\Phi_{cl}(r) \quad \text{and} \quad A_{cl\mu} = -\frac{1}{e}\partial_{\mu}\varphi \mathscr{A}_{cl}(r). \end{equation}
The equations of motion for the profiles $\Phi_{cl}(r)$ and $\mathscr{A}_{cl}(r)$ and other details can be found in the appendix. If we set $m_H = m_V = 1$, which numerically corresponds to substituting $\mu = 1/\sqrt{2}$ and $e = \sqrt{\lambda}$, we find the cosmic string background solution shown in Figure \ref{background}. Since the Higgs mass sets the energy of the theory, all observables should be reasonably smooth functions of $m_V/m_H$. We are always free then to take $m_H = 1$, as we do here. As is well known, $m_V = m_H$ corresponds to the BPS case \cite{Tong}, in which case the $\Phi_{cl}$ and $\mathscr{A}_{cl}$ profiles satisfy first order equations, shown in the appendix. While all the numerical results presented in this paper are for the BPS sector of the theory, we check to ensure that the conclusions are valid in general. That is, we verify that slightly altering $m_V/m_H$ only smoothly alters the results presented below.

To go further, we consider perturbation theory about a stable vacuum with the field fluctuations
\begin{equation} \tilde{\xi}(x) \equiv \phi(x) - \phi_{cl}(x) \quad \text{and} \quad \tilde{\chi}_{\mu}(x) \equiv A_{\mu}(x) - A_{cl\mu}(x). \end{equation}
The linearized equations of motion are then given by the eigenmode equations for the quadratic fluctuation operator $\tilde{\Omega}$, explicitly constructed in \eqref{fluctuation operator}. The ambiguous field normalization is fixed by demanding that their Poisson brackets with their canonical conjugates are each unity. The quantization of the theory is then straightforward by forming the commutators as $i$ times the relevant Dirac brackets \cite{Weinberg}. The only subtlety is that the existence of two zero-modes, calculated explicitly in Section \ref{low-lying}, renders the operator $\tilde{\Omega}$ singular. Physically this corresponds to the fact that the zero-modes are not small fluctuations, i.e. they are massless. In path integral quantization, this problem is remedied by utilizing collective coordinates; for instance, see \cite{Cohn and Periwal} and references therein.

Before considering the fluctuation equations of motion, it is useful to Fourier transform and decompose onto a spin basis. Taking cylindrical coordinates again, we transform the scalar and gauge fields by
\begin{align} \tilde{\xi}(x^0, r, \varphi, x^3) & = \sum_{m \in \mathds{Z}}\int \frac{dp_0 dp_3}{(2\pi)^2} \ \xi_m(p_0, r, p_3)e^{-ip_{\alpha} x^{\alpha}}e^{im\varphi} \\ \tilde{\xi}^*(x^0, r, \varphi, x^3) & = \sum_{m \in \mathds{Z}}\int \frac{dp_0 dp_3}{(2\pi)^2} \ \bar{\xi}_m(p_0, r, p_3)e^{-ip_{\alpha} x^{\alpha}}e^{im\varphi} \\ \tilde{\chi}_{\mu}(x^0, r, \varphi, x^3) & = \sum_{m \in \mathds{Z}}\int \frac{dp_0 dp_3}{(2\pi)^2} \ \chi_{m\mu}(p_0, r, p_3)e^{-ip_{\alpha} x^{\alpha}}e^{im\varphi}, \end{align}
where $\alpha = 0,3$. This immediately gives us the conjugation relations
\begin{equation} \bar{\xi}^*_{-m} = \xi_m \quad \text{and} \quad \chi^*_{(-m)\mu} = \chi_{m\mu}. \end{equation}
Finding the equations of motion in position space, Fourier transforming and then projecting onto the appropriate field basis gives the relationship between the labels of $\xi_m$, $\bar{\xi}_{m'}$ and $\chi_{m''\mu}$ appearing in any equation of motion as
\begin{equation} \{m, m', m''\} = \{m, m-2, m-1\}. \end{equation}
What is the relationship between $m$ and the spin of each mode? The validity of expanding the Fourier fluctuations in angular momentum modes lies in the fact that the direct product of the spacetime symmetry group from breaking down the bulk Poincar\'{e} group and the internal Abelian symmetry group spontaneously breaks down to the diagonal subgroup,
\begin{equation} O_{\text{ext}}(2) \times U_{\text{int}}(1) \longrightarrow U_{\text{diag}}(1). \end{equation}
Here, the generator of this remaining unbroken symmetry is
\begin{equation} J_3 = L_3 - Q, \end{equation}
where $L_3$ is the $O_{\text{ext}}(2)$ generator for rotations about the $z$-axis and $Q$ is the gauge $U_{\text{int}}(1)$ generator (charge operator). Since we are dealing with a charge $1$ string, the actual spin of a given mode is therefore
\begin{equation} \text{spin} = m-1. \end{equation}
Lastly, we need to choose a gauge and an Ansatz for the solutions. We choose the temporal gauge $\chi_0 = 0$ (this is valid for all spin modes). Further, we can always boost a single particle massive state to rest along the string \footnote{This obviously will not capture the massless Nambu-Goldstone modes. These can be exactly calculated via symmetry arguments as given in the following section.}, so we choose the Ansatz $p_3=0$, making the two-dimensional mass $M^2 = p_0^2$. With these conventions, one of the equations of motion is trivially satisfied, leaving four remaining equations. Other than a few general observations, the particular form of these equations is not particularly illuminating, so we write them down completely only in \eqref{eom 1}-\eqref{eom 4}. Note that we work in terms of the one-form components, since geometrically this is the object that must obey the condition of smoothness, not the vector. With our conventions, the $z$-component of the one-form decouples from the other fields. Of course, we can obtain the fields for arbitrary $p_3$ by boosting in the $z$-direction, which will couple $\chi_3$ to the rest of the one-form. 

Each Fourier-transformed field is in general complex, and the form of the equations \eqref{eom 1}-\eqref{eom 4} shows that the solutions decouple. Writing the fields as $\xi_m = \xi_m^r + i\xi_m^i$, the real and imaginary parts of the full equations show that the two decoupled sets of fields are
\begin{equation} \{\xi_m^r, \bar{\xi}_{m-2}^r, \chi_{(m-1)\varphi}^r, \chi_{(m-1)r}^i\} \quad \text{and} \quad \{\xi_m^i, \bar{\xi}_{m-2}^i, \chi_{(m-1)\varphi}^i, \chi_{(m-1)r}^r\}. \end{equation}
The equations show more than just this decoupling; these two sets are degenerate! In fact, it is easily checked that the mode solutions are quite simply related ---
\begin{equation}\label{degenerate} \xi_m^r = \xi_m^i, \quad \bar{\xi}_{m-2}^r = \bar{\xi}_{m-2}^i, \quad \chi_{(m-1)\varphi}^r = \chi_{(m-1)\varphi}^i \quad \text{and} \quad \chi_{(m-1)r}^i = -\chi_{(m-1)r}^r. \end{equation}
This is merely a rotation in internal space. While this result is nontrivial, it could never have been another way since it is due to the unbroken diagonal subgroup symmetry. Since the vacuum retains this residual symmetry, there must exist massive multiplets forming irreducible representations of the diagonal subgroup; this is merely what \eqref{degenerate} is expressing.

We note in passing that different spin sectors (i.e. different choices of $m$) require that certain degrees of freedom vanish for consistency with the equations of motion. While we have found explicit solutions in several of the lower spin sectors, it is not obvious whether physical bound state solutions can be obtained for arbitrarily large spins.

\section{Lowest-Lying Spectrum} \label{low-lying}
\hspace*{1em} We numerically solve the fluctuation equations of motion \eqref{eom 1}-\eqref{eom 4} in Mathematica utilizing a shooting method. We find the modes' Taylor series at the origin utilizing the boundary conditions resulting from smoothness of the scalar and one-form and their Laurent series at spatial infinity, not yet imposing any conditions at infinity. Any one of the parameters in the origin-asymptotic solution can be set to $1$ by linearity. Then we define a function to integrate out the solution from the origin as a function of the initial parameters and match the parameters in the Laurent expansion as a function of the initial parameters; for the numerical solutions we obtain, we find that taking `infinity' to be $r=20$ is more than sufficient for the parameter ranges considered. These parameters are then fixed by FindRoot by imposing the boundary conditions at spatial infinity. We look for localized solutions, but check that in every case we find a continuum of solutions exactly where one would expect; for example, at energies above $m_H$, one can prepare a scalar field plane wave with arbitrary energy $E>m_H$. Each sector can have several bound states below the continuum, or just one or none. To be consistent, all the numerical solutions in the remainder of this paper are obtained for the choice $m_V = m_H=1$, whose background fields are graphed in Figure \ref{background} above. We have performed the same numerical calculations for different values of $m_V/m_H$ and demonstrated that none of the qualitative results depend on the exact ratio as long as we are in the perturbative r\'{e}gime.

\subsection{Nambu-Goldstone Modes}
\hspace{1em} The very presence of the cosmic string implies the existence of Nambu-Goldstone modes. Here, the string breaks translational invariance in the transverse plane, so the pattern of symmetry breaking implies a vector Nambu-Goldstone mode. To be clear, these modes are still scalars on the string worldsheet, but they transform as a vector in the bulk.  Therefore, we expect to find a Nambu-Goldstone massless solution in each of the $m=0$ and $m=2$ sectors. 

Consider an infinitesimal diffeomorphism of the classical soliton solution. The spontaneously broken spacetime symmetries dictate that the zero modes correspond to the variations of the cosmic string along the two independent vector fields whose flows describe the symmetry breaking pattern. In this paper, such a construction is quite simple since a convenient choice of independent vector fields is $\displaystyle Y_1 \equiv \frac{\partial}{\partial x^1}$ and $\displaystyle Y_2 \equiv \frac{\partial}{\partial x^2}$. The variation of the static background with respect to the vector fields is simply its Lie derivative. That is, the two Nambu-Goldstone modes correspond to
\begin{equation} \tilde{\xi} = \pounds_{Y_1}\phi_{cl}, \ \ \tilde{\chi}_{\mu} = \left[\pounds_{Y_1} A_{cl}\right]_{\mu} \quad \text{and} \quad \tilde{\xi} = \pounds_{Y_2}\phi_{cl}, \ \ \tilde{\chi}_{\mu} = \left[\pounds_{Y_2} A_{cl}\right]_{\mu}. \end{equation}
However, in this form, the action of the $U(1)$ is not manifest. To restore this convenience, one must use the \emph{covariant} Lie derivative\footnote{In the geometric viewpoint of the principal $U(1)$-bundle, the Lie derivative achieves covariance when the tangential vector field $\displaystyle \frac{\partial}{\partial x}$ is supplemented by a rotated perpendicular component in the $y$-direction, i.e.~$Y_1 \pm iY_2$. Of course, this method is not needed in the Abelian case, as the final covariant form of the Nambu-Goldstone modes are merely linear combinations of the forms calculated from only the $x$- and $y$-directions, and could be guessed quite easily.}, which in this case is equivalent to using the vector fields $Y_1 \pm i Y_2$. Computing this ordinary Lie derivative in cylindrical coordinates with the top sign gives
\begin{align} \tilde{\xi} & = \frac{e\mu}{\sqrt{\lambda}}\left(\frac{d\Phi_{cl}}{dr}-\frac{1}{r}\Phi_{cl}\right)e^{2i\varphi} \\ \tilde{\chi}_r & = \frac{i}{r^2}\mathscr{A}_{cl} e^{i\varphi} \\ \tilde{\chi}_{\varphi} & = \left(\frac{1}{r}\mathscr{A}_{cl} - \frac{d\mathscr{A}_{cl}}{dr}\right)e^{i\varphi}, \end{align}
while doing so for the bottom sign gives
\begin{align} \tilde{\xi} & = \frac{e\mu}{\sqrt{\lambda}}\left(\frac{d\Phi_{cl}}{dr}+\frac{1}{r}\Phi_{cl}\right) \\ \tilde{\chi}_r & = -\frac{i}{r^2}\mathscr{A}_{cl} e^{-i\varphi} \\ \tilde{\chi}_{\varphi} & = \left(\frac{1}{r}\mathscr{A}_{cl}-\frac{d\mathscr{A}_{cl}}{dr}\right)e^{-i\varphi}. \end{align}

These have precisely the predicted spin properties! In addition, the connection between the real and imaginary components noted in the previous section is manifest. The former expression corresponds to a helicity $+1$ state and the latter to a helicity $-1$ state. In the notation of the previous section, the Nambu-Goldstone solutions are thus
\begin{align}\label{NG up} \text{helicity} \ +1&: \  \  & \xi_2^{NG} = \frac{e\mu}{\sqrt{\lambda}}\left(\frac{d\Phi_{cl}}{dr}-\frac{1}{r}\Phi_{cl}\right) \quad &\chi_{1r}^{NG}  = \frac{i}{r^2}\mathscr{A}_{cl} & \chi_{1\varphi}^{NG} = \frac{1}{r}\mathscr{A}_{cl} - \frac{d\mathscr{A}_{cl}}{dr} \\ \text{helicity} \  -1&: \ & \  \xi_0^{NG}  = \frac{e\mu}{\sqrt{\lambda}}\left(\frac{d\Phi_{cl}}{dr}+\frac{1}{r}\Phi_{cl}\right) \quad & \chi_{-1r}^{NG} = -\frac{i}{r^2}\mathscr{A}_{cl}  &\chi_{-1\varphi}^{NG}  = \frac{1}{r}\mathscr{A}_{cl}-\frac{d\mathscr{A}_{cl}}{dr}. \end{align}

\begin{figure}
\centering
\includegraphics[scale=0.65]{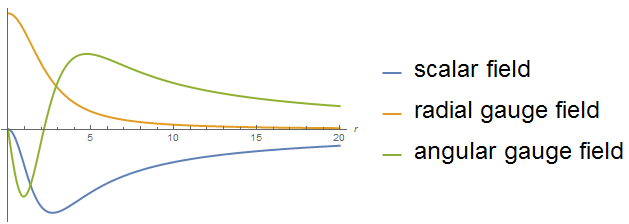}
\caption{Numerical solutions for the helicity $+1$ Nambu-Goldstone mode given exactly by \eqref{NG up}. [Note that we show the modulus of $\chi_{1r}^{NG}$.] Although these solutions might seem to not be sufficiently localized, it is precisely this longer-range behavior which results in `perturbations' around the solution being quite large.}
\label{NG}
\end{figure}

As expected, the gauge field components for both solutions are conjugates of each other while the scalar fields are independent. The positive helicity state is shown in Figure \ref{NG}. Both modes were obtained in Mathematica and shown to exactly obey the second and third equations of motion \eqref{eom 2} and \eqref{eom 3}, since $M_{NG}=0$. While the method to derive these modes by considering the Lie derivative and the principal $G$-bundle over the quotient space seems unnecessary in this simple Abelian case, it is clear that the same method applies to non-Abelian gauge theories. In addition, if one wanted to couple the string to gravity by introducing a curved target space, the above construction is trivial to modify to account for the zero-modes resulting from the Lie derivatives with respect to the appropriate Killing vectors. 

Before moving on, we note the most general form of the Nambu-Goldstone modes, introducing the collective coordinates $X^1$ and $X^2$ which provide the transverse embedding for the cosmic string. Manifest $U(1)$ covariance is no longer necessary, so the general solutions are just the Lie derivatives with respect to the vector fields
\begin{equation} Y_i \equiv \frac{\partial}{\partial x^i} = \frac{\partial r}{\partial x^i}\frac{\partial}{\partial r} + \frac{\partial \varphi}{\partial x^i}\frac{\partial}{\partial \varphi}, \quad i = 1,2, \end{equation}
which have cylindrical components
\begin{equation}\label{components} Y_x^{\mu} = \begin{pmatrix} 0 \\ \cos\varphi \\ \displaystyle -\frac{\sin\varphi}{r} \\ 0 \end{pmatrix} \quad \text{and} \quad Y_y^{\mu} = \begin{pmatrix} 0 \\ \sin\varphi \\ \displaystyle \frac{\cos\varphi}{r} \\ 0 \end{pmatrix}. \end{equation}
Then, the Cartan identity on forms gives us the general solutions
\begin{align}\label{NG1} \tilde{\xi}^{NG} & = X^i Y_i^{\mu}\partial_{\mu}\phi_{cl} \\ \label{NG2} \tilde{\chi}_{\mu}^{NG} & = X^i Y_i^{\nu}\partial_{\nu}A_{cl\mu} + X^i A_{cl\nu}\partial_{\mu} Y_i^{\nu}. \end{align}
The collective coordinates $X^i$ live on the vortex itself. Defining the natural worldsheet coordinates $\sigma^0 = x^0$ and $\sigma^1 = x^3$ (or equivalently the Wick-rotated Euclidean time $\sigma^2 = ix^0$), the collective coordinates are worldsheet scalar fields $X^i(\sigma^0, \sigma^1)$. The equations of motion these fields obey can be inferred directly from compatibility of the general solutions \eqref{NG1} and \eqref{NG2} with the Euler-Lagrange equations of the full theory. It is not difficult to show that this requires the $X^i$ to satisfy free wave equations, which is of course necessary. It is well known that the generalization of the symmetry breaking method of Callan, Coleman, Wess and Zumino (CCWZ) to spacetime symmetries\cite{CCWZ, spacetime1, spacetime2} gives the leading order low-energy action as the Nambu-Goto action, as explained in the appendix. That the effective string theory thus derived leads to a consistent theory was shown in \cite{PS}. The leading corrections to the Nambu-Goto action are then geometric terms such as couplings to the curvature \cite{DFG3}. This effective string theory will reappear in the penultimate section with some interesting consequences once a pseudoscalar is topologically coupled to the string.

\subsection{Least-Massive Mode}
\begin{figure}
\centering
\includegraphics[scale=0.54]{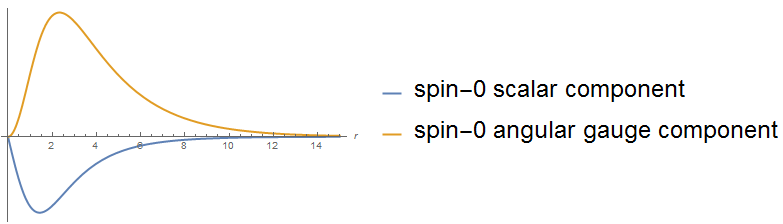}
\caption{The lowest-lying bound state supported by the standalone cosmic string, located in the $m=1$ (spin-$0$) sector. The localization of this solution is clear. Further, it is easily identified with the breather mode \eqref{breather}, and hence is a scalar excitation.}
\label{lowest mode}
\end{figure}

\hspace*{1em} Now we search for the least massive excitation of the string. Na\"{i}vely one would expect this state to exist in the spin zero sector since any spin would cost energy outside the string core for the full field to not be aligned with the background. Indeed, we see the fluctuation equations of motion \eqref{eom 1}-\eqref{eom 4} single out the case $m=1$ profoundly. In this case, several terms vanish from the equations completely, but the largest difference is a shift in the propagating degrees of freedom. Solely when $m=1$ are $\xi_m$ and $\bar{\xi}_{m-2}$ not independent. Actually, it can be shown that these equations are not consistent for all the fields; at the end of the day, only the real field components $\xi_1^r$ and $\chi_{0\varphi}^r$ survive (and of course their degenerate partners). This again makes sense in terms of energy considerations; since the background gauge field is purely angular, one would expect the lowest-lying mode to have only an angular correction without exciting a radial part. The reduced equations of motion for the remaining degrees of freedom (with $m_V = m_H = 1$) are
\begin{align}\label{least eom 1} \left[(2M^2+1-3\Phi_{cl}^2)r^2+2\mathscr{A}_{cl}(2-\mathscr{A}_{cl})-2 \right] \xi_1^r - 2\sqrt{2}\Phi_{cl}(1-\mathscr{A}_{cl})\chi_{0\varphi}^r+2r\frac{d}{dr}\left(r\frac{d\xi_1^r}{dr}\right) & = 0 \\ \label{least eom 2}  2\sqrt{2}r \Phi_{cl}(\mathscr{A}_{cl}-1)\xi_1^r + r(M^2-\Phi_{cl}^2)\chi_{0\varphi}^r-\frac{d\chi_{0\varphi}^r}{dr}+r\frac{d^2\chi_{0\varphi}^r}{dr^2} & = 0.   \end{align}
To illustrate the numerical shooting method, one finds the following Taylor and Laurent series at the origin and infinity ---
\begin{align} &\xi_1^r \underset{r \rightarrow 0}{=} b_1 r + \left(\frac{a_1 b_2}{4\sqrt{2}}-\frac{2M^2+4a_2+1}{16}\right)r^3+\dotsc & \\  &\xi_1^r \underset{r \rightarrow \infty}{=} c_1+\frac{c_2}{r}-\frac{c_1(M^2-1)}{2r^2}-\frac{c_2 M^2}{6r^3}+\dotsc \\ & \chi_{0\varphi}^r \underset{r \rightarrow 0}{=} b_2 r^2 + \left(\frac{a_1 b_1}{2\sqrt{2}}-\frac{b_2 M^2}{8}\right)r^4+\dotsc  \\ & \chi_{0\varphi}^r \underset{r \rightarrow \infty}{=} c_3 + \frac{c_4}{r}-\frac{c_3(M^2-1)}{r^2}-\frac{c_4(M^2-2)}{6r^3}+\dotsc,   \end{align}
where $a_1$ and $a_2$ are two constants that were previously fixed in finding the background solution. After scaling away either $b_1$ or $b_2$ in the origin-asymptotic expansions, we are left with two parameters to shoot with (the remaining expansion parameter and the mass of the mode $M$). As a check, these two starting parameters are set by requiring the constant terms in the infinity asymptotic expansion vanish; this is a consistent system of equations. As mentioned in the Introduction, we fix the fluctuation scale by imposing canonical Poisson brackets of the fields with their conjugates, e.g. $[\tilde{\xi}, \dot{\tilde{\xi}}]_P = 1$. This normalization procedure also illustrates how zero-modes are problematic if one does not use the collective coordinate formalism. However, $M$ is meaningful (and also a highly nontrivial function of the mass scale $m_V/m_H$), and the lowest-lying solution in this spin sector is found to have
\begin{equation}\label{mass} M = 0.881744\dotsc. \end{equation}
The corresponding field configurations are shown in Figure \ref{lowest mode}. This is the lowest-lying mode supported by the cosmic string. Though the above discussion is particularly suggestive in favor of this conclusion, the true test is to show numerically that no lower modes are found in any of the spin sectors surrounding this one. We did this and confirmed that the spin-$0$ state found here indeed has the smallest nonzero mass. For example, the spin-$0$ sector includes several other excited states all with a greater mass.

The question becomes --- is this a scalar or a pseudoscalar excitation? When this is not explicitly indicated by the Lagrangian, the question of transformational behavior of bound states supported by the string is a subtle one, for the string background itself must also be considered. The question here is the behavior under the discrete transformations transverse to the string. The `parity' of a bound state is not merely given by the transformation under spatial reflection, but rather by the transformation under $CP$. Luckily for us, this delicacy can be circumvented by intuition. For a vortex with a spin-$0$ fluctuation, one would expect the least energy to be expended by a small dilatation. That is, it would be sensible for the lowest-lying excitation to be a breather mode. An infinitesimal dilatation corresponds to the Lie derivative with respect to the radial vector field
\begin{equation} \mathcal{X} = r\frac{\partial}{\partial r}. \end{equation}
Since this vector field has no angular component, the Lie derivative of the gauge one-form does not rotate the background field into another direction. Such is necessary for the lowest-lying mode, as argued above. So we wish to know if the solution in Figure \ref{lowest mode} corresponds to the breather mode,
\begin{equation}\label{breather} \xi_1^r \stackrel{?}{=} \pounds_{\mathcal{X}}\phi_{cl} = r\frac{d\phi_{cl}}{dr} \quad \text{and} \quad \chi_{0\varphi}^r \stackrel{?}{=} \left[\pounds_{\mathcal{X}} A_{cl}\right]_{\varphi} = r\frac{dA_{cl\varphi}}{dr}. \end{equation}
Substituting these expressions into the spin-$0$ sector equations \eqref{least eom 1} and \eqref{least eom 2}, we find that the equalities hold, and hence the lowest-lying excitation of Figure \ref{lowest mode} is indeed the breather mode. This is an explicit confirmation that our perturbation theory is valid. Therefore, \emph{the cosmic string's lowest-lying mode is a scalar breather excitation.} While this is precisely the answer one might expect, we shall show in the next section that it is not entirely trivial. 

The remainder of the spectrum can be built up in exactly the same fashion. For each sector, one utilizes the shooting method to match asymptotic solutions until the solutions become nonlocalized, i.e. when one reaches the continuum. One must still be careful, however, in properly accounting for the degrees of freedom. An arbitrary spin sector is coupled to its negative sector due to the nature of our projection, i.e. the equations \eqref{eom 1}-\eqref{eom 4} for $m$ and $2-m$ have the \emph{same} field content. Thus what appears at first glance to be an underdetermined system now appears to be overdetermined since we have four complex fields governed by six complex equations. Such a system is in general inconsistent, but all is not lost. The structure of the equations can allow for the fields and their equations to be divided into two separate systems, one of which is consistent and the other inconsistent. This is precisely what happens in the spin-$0$ system, for example. The Nambu-Goldstone modes from the previous subsection also result from a consistent system because the first equation of motion \eqref{eom 1} is only valid for nonzero mass $M$ (the entire pre-Fourier-transformed equation is acted upon by a time derivative), and the doubling of the helicity $\pm 1$ sectors then gives four equations for the four fields, $\xi_2$, $\xi_0$, $\chi_{1\varphi}$ and $\chi_{1r}$. 

We have not yet had much to say about the fourth equation of motion \eqref{eom 4} which decouples $\tilde{\chi}_{(m-1)3}$ from the remainder of the fields. Incidentally, that equation does possess a mathematical `bound state' solution that is degenerate with the breather mode just found, but this is not physically interesting itself. This field component essentially does not interact with anything else in this theory (due to the nature of the perturbation theory), and can be momentarily ignored. What \emph{is} important, however, is the fact that the final equation \eqref{eom 4} has a solution degenerate with the least massive excitation above. It shall play an important role in the next section when the theory is expanded to include other interactions.

\section{Scalars and Pseudoscalars; Coupling to the String}\label{externals}
\subsection{External Spin-0 Field}
\hspace*{1em} We have shown that a pure Abelian vortex has a lowest-lying scalar mode of mass $M = 0.881744\dotsc$. In this section we shall briefly explore a simple extension to the theory by adding an additional spin-$0$ field that interacts with the Higgs field. We are interested in whether or not we can change the properties of the least massive excitation. In addition to its mass, we are primarily interested in its parity. 

Let us couple the theory to a real external spin-0 field, $\sigma$, by adding to the Abelian Higgs model Lagrangian the terms\footnote{We do not add a mass term for $\sigma$, but it would be trivial to do so as it would just shift the mass of the bound state found below.}
\begin{equation}\label{modified 1} \mathscr{L} \supset \frac{1}{2}\partial_{\mu} \sigma \partial^{\mu} \sigma - \beta \lambda \phi^* \phi \sigma^2. \end{equation}
The factor $\lambda$ is explicitly written in the quartic interaction so that $\beta$ itself is the physical coupling. We na\"{i}vely expect $\sigma$ to have a localized bound state since its mass (squared) changes continuously from $0$ at the string core to $\beta m_H^2$ at transverse spatial infinity (this is why the interaction coupling is not divided by $2$). The modification this interaction makes to our perturbation theory is nearly trivial. The linearized equations of motion \eqref{eom 1}-\eqref{eom 4} are unchanged, since $\sigma$ enters quadratically. Further, only the background Higgs field enters in the $\sigma$ equation of motion for the same reason. Thus, we need only solve a modified Klein-Gordon equation. Performing the same Fourier transform and spectral decomposition
\begin{equation} \sigma = \sum_{m \in \mathds{Z}}\int \frac{dp_0}{2\pi}\sigma_m e^{-ip_0 x^0}e^{im\varphi}, \end{equation}
which also yields the reality constraint $\sigma_{-m}^* = \sigma_m$, the resulting equation we must solve is
\begin{equation} \frac{1}{r}\frac{d}{dr}\left(r\frac{d\sigma_m}{dr}\right)+\left(M_{\sigma}^2 -\frac{m^2}{r^2}-2\beta \lambda \phi_{cl}^*\phi_{cl}\right)\sigma_m = 0. \end{equation}
Even though we shall still have to resort to a numerical solution, several properties are obvious just from looking at this equation. First, the leading order expansion at the origin must be $\sigma_m \propto r^m + \dotsc$ and hence the only possible bound states are in the $m=0$ sector ($m$ is the spin in this case because the internal symmetry generator is missing). This result is intuitive; the behavior is very familiar from quantum mechanics, and a bound state must interact directly with the string core. Secondly, the exponential solution at infinity is
\begin{equation} \sigma_m(r) = c_1 e^{-\sqrt{\beta m_H^2 - M_{\sigma}^2}r} + c_2 e^{\sqrt{\beta m_H^2 - M_{\sigma}^2}r}. \end{equation}
Therefore, the continuum in each sector begins at $M_{\sigma} = \sqrt{\beta}m_H$, which makes sense physically since it is the real scalar mass at transverse spatial infinity. Both of these observations are confirmed numerically. 

\begin{figure}[t]
\centering
\includegraphics[scale=0.65]{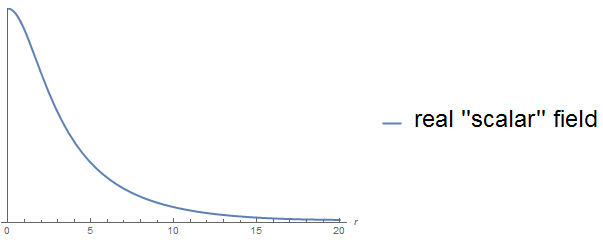}
\caption{The single bound state for an external real field coupled to the cosmic string via a quartic interaction, which is located in the spin-$0$ sector. When the positive coupling satisfies $\beta < 1$, this mode becomes the lowest-lying excitation. However, there is no interaction that unambiguously dictates its parity.}
\label{scalar}
\end{figure}

In solving the equation, choose $\beta = \frac{1}{2}$ for concreteness. We find that there exists a unique\footnote{Unique only up to the addition of an arbitrary mass term of no consequence to us.} bound state of the real scalar with $m=0$ and mass $M_{\sigma} = 0.673424\dotsc$; this is lower than the excitation produced by the string alone! The mode is shown in Figure \ref{scalar}. Therefore, in the modified theory \eqref{modified 1}, the lowest-lying mode is this $\sigma_0$. Indeed, this is true for quartic coupling $\beta<1$. For $\beta = 1$, $\sigma_0$ is degenerate with the breather in the previous section. For $\beta > 1$, the bound state still exists but with a mass greater than that of the breather. 

Is this spin-$0$ mode a scalar or a pseudoscalar? This is an ambiguous question! To an extent, the parity operator in a theory can be redefined in several ways to shuffle around phases. However, we have no such freedom here because the existing string background and its corresponding particle excitations unambiguously determine their parity (which, as stated before, is really their transformations under $CP$). Since no redefinition can be made at this point, the parity of any external fields added to the theory must be determined from their interactions. But \eqref{modified 1} provides no information to this end. To rectify this, we could add another interaction that forces $\sigma$ to be a scalar field proper. But let us delve immediately into the interesting case by forcing $\sigma$ to be a pseudoscalar. 

\subsection{External Pseudoscalar and Spectrum Shift}
\hspace*{1em} Let us add an interaction term linear in $\sigma$, coupling it to a bosonic term odd under $CP$. This construction will give us a pseudoscalar $\sigma$ by parity invariance. We want the lowest dimension operator possible so it appears in the low-energy effective string theory (more on this in the next section). Thus we cannot construct it using the Higgs field $\phi$. Then, there is essentially a unique operator of the gauge forms with the correct transformational properties, $F \wedge F$. Therefore, consider the new theory defined by\footnote{It should be emphasized that we are not considering the most general renormalizable interactions. In principle we could also have a quartic $\sigma^4$ interaction, for example.}
\begin{equation}\label{modified 2} \mathscr{L} = -\frac{1}{4}F_{\mu \nu}F^{\mu \nu} + \nabla_{\mu}\phi^* \nabla^{\mu}\phi +\frac{1}{2}\partial_{\mu}\sigma \partial^{\mu}\sigma -\frac{\lambda}{2}\left(\phi^* \phi - \frac{\mu^2}{\lambda}\right)^2 - \beta \lambda \phi^* \phi \sigma^2 - \alpha e \sigma F \wedge F. \end{equation}
For the moment we abuse notation by writing the new interaction as a four-form to emphasize its geometrical underpinnings. This $\sigma F \wedge F$ interaction now makes $\sigma$ unstable, as it can always decay at least to two Nambu-Goldstone bosons for any value of the coupling $\alpha$, an important fact for later considering the effective string theory.

This case is also not too much harder to handle than in the absence of this interaction. We see that as far as our linearized perturbation theory is concerned, $\sigma$ is coupled only to $\tilde{\chi}_3$. But $\tilde{\chi}_3$ decoupled earlier in \eqref{eom 4}, so we only have to derive and solve two equations of motion. This is a powerful consequence of our gauge and Ans\"{a}tze. Computing the equations and performing the transforms and projections yield
\begin{align} \frac{1}{r}\frac{d}{dr}\left(r\frac{d\sigma_m^r}{dr}\right)+\left(M_{\sigma}^2-\frac{m^2}{r^2}-2\beta \lambda \phi_{cl}^*\phi_{cl}\right)\sigma_m^r + \frac{2\alpha e M_{\sigma}}{r}\frac{dA_{cl\varphi}}{dr}\chi_{m3}^i & = 0 \\ \frac{1}{r}\frac{d}{dr}\left(r\frac{d\chi_{m3}^i}{dr}\right) + \left(M_{\sigma}^2-\frac{m^2}{r^2}-2e^2 \phi_{cl}^* \phi_{cl}\right)\chi_{m3}^i + \frac{2\alpha e M_{\sigma}}{r}\frac{dA_{cl\varphi}}{dr}\sigma_m^r & = 0. \end{align}

\begin{figure}
\centering
\includegraphics[scale=0.63]{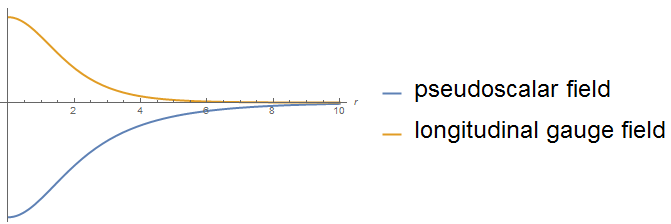}
\caption{The least massive pseudoscalar bound state due to the $\sigma F \wedge F$ interaction, again in the spin-$0$ sector. For all coupling constants, the mass of this bound state is below that of the pure string lowest excitation. Therefore in this theory, we have a pseudoscalar least massive excitation.}
\label{pseudoscalar}
\end{figure}

These equations become mirrored for the special choice $m_V/m_H = \beta$. If we choose $\beta = 1$, the entire problem has an enhanced symmetry. This is the case in which the background fields obey the first order BPS equations which saturate the Bogomol'nyi bound as well as the pseudoscalar and gauge form $z$-component possess an interchange symmetry. Once again, we shall explicitly present the profile solutions with the $\beta = \frac{1}{2}$ from the previous subsection, still taking $\alpha$ to be unity. Doing this indeed yields a bound state in the $m = 0$ sector with a mass of $M_{\sigma} = 0.575467\dotsc$. The corresponding solutions are shown in Figure \ref{pseudoscalar}. Once again, we find that this massive state is below the first excited state produced by the string alone. This time, however, we know it corresponds to a pseudoscalar. \emph{One can produce lowest-lying pseudoscalar states with pure bosonic field content.} The presence of the $\sigma F \wedge F$ interaction shifts the cosmic string spectrum to now start with a pseudoscalar mode instead of a scalar one. It is not surprising itself that bosonic soliton systems can support pseudoscalar modes, but it seems nontrivial that these modes can naturally be excited before ones with more symmetric origins, for instance the breather. 

\begin{figure}
\centering
\includegraphics[scale=0.46]{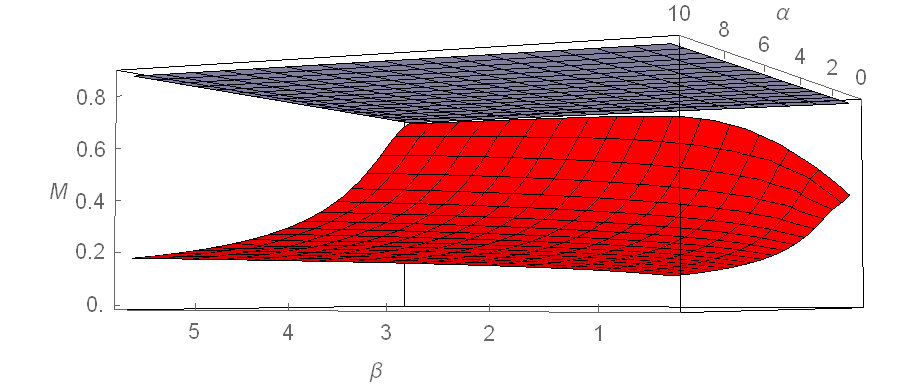}
\caption{The mass (red) of the lightest pseudoscalar mode for the BPS sector as a function of the two couplings, $\beta$ and $\alpha$, defined in \eqref{modified 2}. The plane (blue) at the breather mode's mass is also included. For small $\alpha$ and large $\beta$, the pseudoscalar's mass approaches that of the breather asymptotically.}
\label{pretty graph}
\end{figure}

In fact, the result here is slightly more general. The profiles in Figure \ref{pseudoscalar} are for only a specific value of the couplings $\beta$ and $\alpha$. As mentioned in the previous subsection, setting $\alpha=0$ and varying $\beta>0$ shows that the least massive mode for the spin-$0$ field $\sigma$ can be either more or less massive than the cosmic string's breather. However, as soon as a nonzero $\alpha$ is introduced, the pseudoscalar mass $M_{\sigma}$ immediately becomes the least massive. For a given region of the $(\beta, \alpha)$ parameter space, the least massive $\sigma_0$ excitation as a function of the parameters, $M_{\sigma}(\beta,\alpha)$, asymptotically approaches the mass of the breather mode found above. Hence this mass function appears to have a discontinuity as $\alpha$ approaches zero, since we have demonstrated that when $\alpha=0$, $M_{\sigma}$ can exceed the mode's mass. This state of affairs is shown in our BPS case in Figure \ref{pretty graph}, which is an interpolation of a finely meshed $(\alpha, \beta)$ grid where at each point the bound state mass $M_{\sigma}$ was numerically determined. In this case, the asymptotic r\'{e}gime is given by\footnote{Note that we are indeed justified in allowing $\beta$ and $\alpha$ to become significantly larger than $1$ (in our units defined by $m_H = m_V = 1$), since perturbation theory is defined in terms of $\beta \lambda$ and $M_5 \alpha e$, where $M_5$ is some mass scale introduced for the dimension $5$ operator. We are then free to choose $e,\lambda \ll 1$, $e^2 \sim \lambda$ such that these are small. Specifically, the larger the range of $(\alpha, \beta)$ space we include, the smaller the domain of validity of perturbation theory.}
\begin{equation} \lim_{\substack{\alpha \rightarrow 0 \\ \beta \gg 1}}M_{\sigma}(\beta, \alpha) = M_{\text{breather}} \approx 0.881744\dotsc. \end{equation}
This is a surprising result. For small but nonzero $\alpha$ the mass of the pseudoscalar mode stops tracking the mass of the spin-0 breather mode found above, i.e.~for the case $\alpha=0$ exactly. In the next section we shall consider the effective string theory consisting of the massless degrees of freedom. The mass of the lightest excitation sets the scale at which this effective theory breaks down, since the tension of the string is much larger than the square of the breather mass. Since $\sigma F \wedge F$ is a dimension $5$ operator, we would like to say that $\alpha$ is essentially an inverse mass. Thus the limit $\alpha \rightarrow 0$ would be equivalent to sending the corresponding mass scale to infinity. So we see that the scale at which the massive string excitations are no longer irrelevant is not the scale provided by the coupling $\alpha$. It might seem puzzling at first that the most `relevant' terms to include in an effective string theory are the ones ultimately deriving from an irrelevant dimension $5$ operator, even in the limit when this operator's scale blows up. In some regards, what appears to be a dimension $5$ operator behaves like a lower dimensional operator. Note that this is distinct from the concept of large anomalous dimensions which arise in strongly coupled theories which also lead to the failure of na\"{i}ve dimensional analysis; here it is the presence of field vevs which skew our analysis. This is clarified below by an explicit calculation in the effective theory showing that our dimension $5$ operator produces a renormalizable term, to which we now turn.

\section{Pseudoscalars in Effective String Theory} \label{effective string theory}
\hspace*{1em} The pseudoscalar introduced in the previous section turns out to have more interesting properties. Let us now eliminate its quartic coupling to the Higgs field, so that its interactions are completely governed by the Lagrangian four-form term $\mathscr{L}_I = -\alpha \sigma F \wedge F$. Specifically, removing the $\beta$ interaction term reinstates the shift symmetry of $\sigma$ that in QCD is used to describe it as the Goldstone boson for the broken Peccei-Quinn symmetry \cite{Peccei-Quinn}. In the following two subsections, we show that this simple topological interaction produces not only an interaction of the string with a coherent Kalb-Ramond background, but also yields a worldsheet axion interaction at one-loop in the matching to the effective string theory. To be explicit with our conventions, we introduce a local orthonormal frame basis and its corresponding vielbein $e^{\mu}_a$ such that the Levi-Civita symbols read
\begin{align} \epsilon_{\mu_0 \cdots \mu_{d-1}} & = \frac{1}{\sqrt{-g}}\epsilon_{a_0 \cdots a_{d-1}}e^{a_0}_{\mu_0} \cdots e^{a_{d-1}}_{\mu_{d-1}} \\ \epsilon^{\mu_0 \cdots \mu_{d-1}} & = \sqrt{-g} \epsilon^{a_0 \cdots a_{d-1}} e^{\mu_0}_{a_0} \cdots e^{\mu_{d-1}}_{a_{d-1}}, \end{align}
with $\epsilon_{0123} = -\epsilon^{0123}=1$. In these conventions then, the covariant symbol is \\ $\epsilon_{a_0 a_1 a_2 a_3} e^{a_0}_{\mu_0} e^{a_1}_{\mu_1} e^{a_2}_{\mu_2} e^{a_3}_{\mu_3} = \sqrt{-g}\epsilon_{\mu_0 \mu_1 \mu_2 \mu_3}$ and hence the pseudoscalar interaction's contribution to the action is
\begin{equation} S \supset -\alpha \int \sigma F \wedge F = -\frac{\alpha}{4}\int d^4 x \ \sigma F_{\mu \nu}\epsilon^{\mu \nu \lambda \rho}F_{\lambda \rho}. \end{equation}

\subsection{Coherent Kalb-Ramond Background}
\hspace*{1em} We currently have two types of massless degrees of freedom which determine the low-energy effective theory; the Nambu-Goldstone modes $X^i$ from broken translational invariance and the newly introduced $\sigma$ field. While the standard CCWZ coset construction restricts the interactions of $X^i$ to the Nambu-Goto action plus a tower of geometric invariants \cite{DFG3}, it provides little useful information about the coupling of the pseudoscalar $\sigma$ since it transforms trivially under the unbroken string $ISO(1,1)$. Thus to see how $\sigma$ appears in the low-energy theory, we must explicitly perform the matching to the effective string theory. This is simple at tree-level --- we merely see what form $\sigma F \wedge F$ takes for the Nambu-Goldstone modes $A_{\mu} = \tilde{\chi}_{\mu}^{NG}$. It turns out that we can eliminate $\sigma$ entirely from the theory by exchanging it for a two-form. Since $F \wedge F$ is the exterior derivative of the Chern-Simons three-form, only the derivative of $\sigma$ appears significant in the theory. Specifically, we equate the Hodge dual of the pseudoscalar's exterior derivative with an exact three-form, i.e.
\begin{equation}\label{dual} \ast d\sigma = H = dB. \end{equation}
The kinetic term for the pseudoscalar then turns into a kinetic piece $H_{\mu \nu \lambda}H^{\mu \nu \lambda}$. Of course, we must keep the degrees of freedom the same, so the two-form gauge field $B_{\mu \nu}$ can only have one degree of freedom. This just necessitates a choice of gauge when the field strength $H$ drastically increased the gauge degrees of freedom; we shall not write this out explicitly, but it should be emphasized that it has been done. Now the interaction term in the action reads
\begin{equation}\label{alternate} S \supset \alpha \int d^4 x \left[A^b F^{cd} + A^c F^{db} + A^d F^{bc}\right]\left[\partial_b B_{cd} + \partial_c B_{db} + \partial_d B_{cb}\right]. \end{equation}
We shall now show that this $B_{\mu \nu}$ corresponds to a bulk Kalb-Ramond field, at least at tree level. To see how this plays out explicitly for an Abelian string in $(3+1)$-dimensions, we pull out the completely massless interaction
\begin{equation} S \supset -\alpha \int \sigma F \wedge F \supset -\alpha \int \sigma d\tilde{\chi}^{NG}\wedge d\tilde{\chi}^{NG}, \end{equation}
where from \eqref{NG2} the Nambu-Goldstone one-form components are
\begin{equation} \tilde{\chi}^{NG}_{\mu} = X^i Y_i^{\nu}\partial_{\nu}A_{cl\mu}+X^i A_{cl\nu}\partial_{\mu}Y^{\nu}_i, \end{equation}
and the cylindrical vector components are given explicitly in \eqref{components} for the specific Higgs model considered in this paper. We denote by beginning-alphabet Greek letters the two worldsheet coordinates and by mid-alphabet Latin letters the two transverse coordinates. By using the explicit representation given in \eqref{components} and the fact that the only nonvanishing background gauge profile component is $A_{cl\varphi}$, it is straightforward to show that the tree-level part of the action for the Nambu-Goldstone modes becomes
\begin{multline} \int \sigma F^{NG}\wedge F^{NG} \\ = -\int d^4 x \ \sigma \epsilon^{\alpha \beta m n}\left[\left(Y_i^r \partial_r A_{cl,m} + A_{cl,r}\partial_m Y_i^r\right)\left(Y_j^s \partial_s A_{cl,n} + A_{cl,s}\partial_n Y_j^s\right)\right]\partial_{\alpha}X^i \partial_{\beta} X^j. \end{multline}
Therefore, we find that Abelian string theories with an external pseudoscalar interacting via the topological term $F \wedge F$ have in their effective string theory the interaction
\begin{equation}\label{Kalb-Ramond} S \supset \int \sigma F \wedge F \supset -\frac{1}{4\pi\alpha'}\int_{\Sigma}d^2 \sigma \ \epsilon^{\alpha \beta}\partial_{\alpha}X^i \partial_{\beta}X^j B_{ij}, \end{equation}
where 
\begin{multline} B_{ij}(\tau,\sigma) = 4\pi\alpha' \int d^2 x_{\perp} \ \sigma \epsilon^{mn}\left[\partial_r A_{cl,m}\partial_s A_{cl,n}Y^r_{[i}Y^s_{j]} \right. \\ \left. + 2A_{cl,r}\partial_s A_{cl,n}\left(\partial_m Y^r_{[i}\right)Y^s_{j]} + A_{cl,r}A_{cl,s}\partial_m Y^r_{[i}\partial_{|n|}Y^s_{j]}\right], \end{multline}
where the integral is over the transverse coordinates and indices between square brackets are to be antisymmetrized. This is the complete tree-level matching to the effective theory, and it precisely takes the form of a Kalb-Ramond interaction \cite{Polchinski}! In this respect, the pseudoscalar $\sigma$ at tree-level produces a background Kalb-Ramond field coherently coupled to the string. This is an outstanding result --- it appears as if an irrelevant operator in the UV theory has given rise to a relevant (or at least marginal) operator in the effective theory. As mentioned above, this behavior ultimately arises from the way vacuum expectation values alter the power counting for operators. The fact that this tree-level matching of the dimension $5$ operator yields a renormalizable interaction explains why the appropriate mass scale to consider is not $\propto 1/\alpha$. To be sure, the coupling $\alpha$ still determines the strength of the effective string interaction, but not directly the scale at which that operator is relevant. 

Specializing to the case of the Abelian Higgs model considered here, the pattern of symmetry breaking shows that the $-\alpha \sigma F \wedge F$ interaction appears as the same interaction as \eqref{Kalb-Ramond}, where the Kalb-Ramond field is given by
\begin{equation}\label{Abelian KR} B_{ij}(\tau,\sigma) = 4\pi \alpha' \alpha \epsilon_{ij}\int dr d\varphi \ \sigma \left[\frac{2}{r^2}A_{cl\varphi}\frac{dA_{cl\varphi}}{dr}-\frac{\left(A_{cl\varphi}\right)^2}{r^3}\right]. \end{equation}
The only dependence on the worldsheet coordinates is due to $\sigma$ itself. Furthermore, the interaction dynamics in the quantum string theory are governed entirely by $\sigma$, reaffirming the interchangeability of the equivalence of the pseudoscalar and the two-form. In fact, the expression \eqref{Abelian KR} may be viewed as gauge fixing the general relation \eqref{dual} to eliminate the spurious degrees of freedom from introducing the two-form gauge field.

\subsection{Emergence of a Worldsheet Axion Interation}
\begin{figure}
\centering
\includegraphics[scale=0.45]{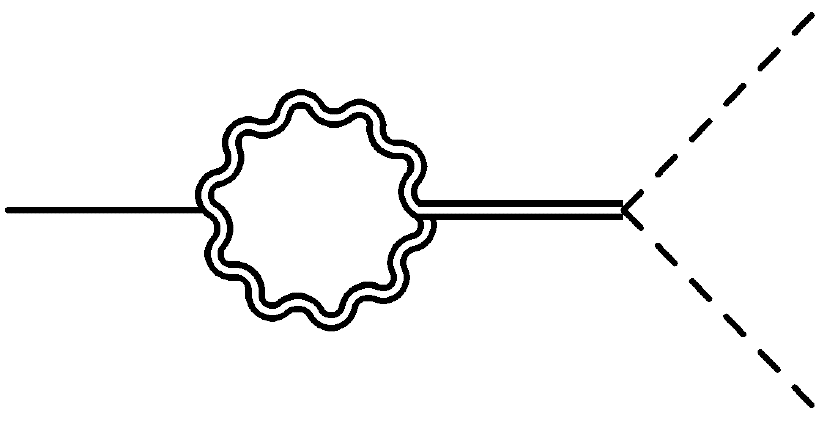}
\caption{A one-loop diagram in the UV which will contribute to the pseudoscalar decaying into two worldsheet scalars; see the text for the line notation.}
\label{loop feynman}
\end{figure}

\hspace*{1em} The appearance of the Kalb-Ramond field \eqref{Kalb-Ramond} in the effective string theory is the result of tree-level matching. What happens when one matches at one-loop order? Specifically, let us focus on the lowest dimension operators emergent from one-loop calculations, i.e. to second order in the $X^i$ fields. That is, we want the amplitude for the pseudoscalar $\sigma$ to decay into $X^i + X^j$. Since the Goldstones are derivatively coupled, decay into more Goldstones is suppressed by the appropriate powers of momenta. The matching performed in the previous subsection corresponds to the tree level diagram of $\sigma$ decaying directly into an $X^i$ and an $X^j$, which appears unmodified in the effective string theory as the interaction $\epsilon^{\alpha \beta}\partial_{\alpha}X^i \partial_{\beta}X^j B_{ij}$. At one loop we encounter graphs with the UV massive particles. Denote the massless $\sigma$ field by a solid line, a worldsheet scalar $X^i$ by a dashed line,  a massive $\sigma$ excitation by a doubled straight line and a massive $A_{\mu}$ gauge excitation by a doubled wavy line. Then a graph like that shown in Figure \ref{loop feynman} will also contribute to the lowest order $\sigma \rightarrow X^i + X^j$ decay of interest. 


At higher order we have the obvious extensions of the above diagrams, namely the repetition of the loop shown in Figure \ref{loop feynman}, the insertion of an $X^i$ in a given loop and the addition of worldsheet scalars as external legs for more general decays. These latter two contributions are higher order in the Goldstones for any given loop order. It is only when we add up all these contributions that the resulting terms in the effective string theory action are invariant under the nonlinearly realized Lorentz transformations, i.e. under
\begin{equation}\label{nonlinear} \delta_{\epsilon}^{\alpha i} X^j = \epsilon\left(\delta^{ij}\sigma^{\alpha}+X^i \partial^{\alpha}X^j\right), \ \delta_{\epsilon}^{\alpha i}\sigma = 0, \end{equation}
which corresponds to the broken boost/rotation generators $J_{\alpha i}$ \cite{nonlinear realization}. Let us now calculate the one-loop diagram of Figure \ref{loop feynman} and match to the effective theory\footnote{In fact, for this decay process, the higher loop diagrams such as the one shown in the main text essentially factorize, meaning that we \emph{only} have to match up to one-loop.}. Recall that the $\sigma F \wedge F$ interaction means that the pseudoscalar interacts with the Goldstones via $B\sigma \epsilon_{ij}\epsilon^{\alpha \beta}\partial_{\alpha}X^i \partial_{\beta} X^j$ and interacts with the massive gauge excitations as $C\sigma \epsilon^{\mu \nu \lambda \rho}\partial_{\mu}\tilde{\chi}_{\nu}\partial_{\lambda}\tilde{\chi}_{\rho}$, where $B$ and $C$ are constants (though $B$ does depend on the background gauge field profile). With these conventions, this one-loop diagram is

\begin{multline}
\text{diagram}_{ij} = \frac{BC^2 \eta^{\mu \sigma}\eta^{\lambda \kappa}\epsilon_{\sigma \nu \kappa \rho}\epsilon^{\bar{\mu}\nu \bar{\lambda}\rho}\epsilon_{ij}\epsilon^{\alpha \beta} k_{1\alpha} k_{2\beta}}{\widetilde{M}_{\sigma}^2} \\ \times  \int_0^1 \hspace{-0.5ex} dx \int \hspace{-0.3ex}\frac{d^4 q}{(2\pi)^4}\frac{2x(1-x)p_{\bar{\mu}}q_{\mu}[p_{\bar{\lambda}}q_{\lambda}+p_{\lambda}q_{\bar{\lambda}}]-(2x^2-2x+1)p_{\lambda}p_{\bar{\lambda}}q_{\mu}q_{\bar{\mu}}}{[q^2-(\widetilde{M}_{\chi}^2-x(1-x)p^2-i\epsilon)]^2},  \end{multline}
where $\widetilde{M}_{\chi}$ and $\widetilde{M}_{\sigma}$ are the masses of the gauge and scalar excitations which are to be integrated out and $p_{\mu}$ and $k_{i\alpha}$ are the momenta of the pseudoscalar and Nambu-Goldstone bosons, respectively. Note that both $\sigma$ and the $X^i$ have worldsheet momenta in the above diagram, but we integrate over all four-dimensional momenta for the massive particles living in the bulk. We are interested in the lowest derivative-order contribution; note that the terms with zero loop momenta and four loop momenta in the numerator both vanish by antisymmetry, leaving only the quadratic pieces above. Then, using the fact that $p_{\alpha}=k_{1\alpha}+k_{2\alpha}$ but keeping $\sigma$ off-shell (so that $p^2 = 2k_1 \cdot k_2 \neq 0$), it is straightforward to see that the lowest order contribution, ignoring the $\frac{1}{\epsilon}-\gamma+\ln 4\pi$ piece \`{a} la $\overline{MS}$, from the one-loop correlator is
\begin{equation} \text{diagram}_{ij} \supset \frac{3iBC^2 \widetilde{M}_{\chi}^2}{2\pi^2 \widetilde{M}_{\sigma}^2}\epsilon_{ij}\epsilon^{\alpha \beta}k_{1\gamma}k_{1\alpha}k_2^{\gamma}k_{2\beta}. \end{equation}
This is precisely the tree-level contribution one gets from the interaction in the effective 2D Lagrangian
\begin{equation}\label{worldsheet axion} \mathscr{L}_I^{\text{eff}} \supset \frac{\alpha}{8\pi^2}\sigma \epsilon_{ij}\epsilon^{\alpha \beta} \partial_{\alpha}\partial_{\gamma}X^i \partial_{\beta}\partial^{\gamma}X^j, \end{equation}
where\footnote{There are other contributions at one-loop; they all have the same structure and only modify the precise coefficient here.}
\begin{equation} \alpha = 12BC^2 \left(\frac{\widetilde{M}_{\chi}}{\widetilde{M}_{\sigma}}\right)^2 \end{equation}
(not to be confused with what was called $\alpha$ earlier, which is now tucked away in the constants $B$ and $C$). This looks exactly like the leading order contribution to the coupling of $\sigma$ to the extrinsic curvature of the string. In the literature, the string self-intersection number interaction
\begin{equation} \mathscr{L}_I = \frac{\alpha}{8\pi^2}\phi \epsilon_{ij}\epsilon^{\alpha \beta}K_{\alpha \gamma}^{\phantom{\alpha \gamma}i}K_{\beta}^{\phantom{\beta}\gamma j}, \end{equation}
with extrinsic curvature
\begin{equation} K_{\alpha \gamma}^{\phantom{\alpha \gamma}i} = \nabla_{\alpha}\partial_{\gamma}X^i = \partial_{\alpha}\partial_{\gamma}X^i + \mathcal{O}(X^3), \end{equation}
is said to couple the worldsheet scalars $X^i$ to the \emph{worldsheet axion} $\phi$ \cite{DFG1,DFG2,axion}. The first subleading term for our pseudoscalar $\sigma$ in the effective string theory appears identical to that of a worldsheet axion. The interaction term \eqref{worldsheet axion} is not invariant under the non-linearly realized Lorentz symmetries \eqref{nonlinear}, but this is a necessary condition for all terms in an effective string action \cite{nonlinear realization}. To be invariant, it must be supplemented by additional terms higher order in the Goldstone bosons. These are precisely provided by the Levi-Civita connection in the covariant derivative part of the extrinsic curvature. Physically they arise from the additional matching terms mentioned above, namely when extra Nambu-Goldstone bosons are inserted either in loops or as external legs.

\section{Conclusions}
\hspace*{1em} In this work, we have presented an analysis of the cosmic string spectrum, paying particular attention to the lowest-lying modes. We explicitly calculated the Nambu-Goldstone modes and then showed that the lowest-lying excitation of the string itself is the scalar breather mode. Along the way, we have emphasized the origins of all the spectrum properties in terms of the pattern of symmetry breaking and the residual invariances left over from such symmetry breaking. 

By coupling the string to an external spin-$0$ field via a quartic interaction supplemented by a $\sigma F \wedge F$ interaction which forces $\sigma$ to be a pseudoscalar, the lowest-lying excitation becomes the pseudoscalar state for all choices of the quartic and four-form couplings. When only the topological $\sigma F \wedge F$ interaction is present, the pseudoscalar appears in the effective string theory as a coherent Kalb-Ramond background at tree-level and as a worldsheet axion interaction at one-loop in the matching.

The methods used here to study Abelian vortices generalize to the perturbation theory of any topological soliton in a theory which does not feature confinement. Unfortunately, the interesting case of QCD flux tubes are beyond the scope of this treatment. The recent developments of calculations based on approximate integrability which agree well with existing lattice data strongly suggest the existence of the light worldsheet axion \cite{DFG1, axion}. This still needs to be understood better, and if this axion indeed exists, one question is to what extent it appears in other theories. For example, it appears to exist in $(3+1)$-dimensional gluodynamics, but not in $(2+1)$-dimensional gluodynamics. This result also appears here, where the worldsheet axion which appears in the bulk via a Chern-Simons type interaction is only present in $d=4$. It will be interesting to see what becomes of this axion and of the nature of pseudoscalars in general gauge theories.

\appendix
\section{Abelian Vortex and Equations of Motion}
In this appendix, some details not crucial to the text are presented which are useful for reproducing the above calculations. To be as self-contained as possible, we also very briefly review the theory of a classical Abelian vortex for the reader less familiar with topological solitons. A nice introductory account is given in \cite{Rubakov}. 

We consider here a vortex in a ($3+1$)-dimensional $U(1)$ gauge model, the field content being a complex scalar $\phi$ and a vector field $A_{\mu}$. The standard Lagrangian is given by
\begin{equation}\label{Lagrangian} \mathscr{L} = -\frac{1}{4}F_{\mu \nu}F^{\mu \nu} + \nabla_{\mu}\phi^* \nabla^{\mu}\phi - \frac{\lambda}{2}\left(\phi^* \phi - \frac{\mu^2}{\lambda}\right)^2 \quad \text{with} \quad \nabla_{\mu} = \partial_{\mu}+ieA_{\mu}. \end{equation}
With these conventions, the theory has the local symmetry 
\begin{equation} \phi(x) \rightarrow e^{-i\alpha(x)}\phi(x) \quad \text{and} \quad A_{\mu}(x) \rightarrow A_{\mu}(x) + \frac{1}{e}\partial_{\mu}\alpha(x). \end{equation}
The nonvanishing vacuum expectation value of $\phi$ spontaneously breaks the $U(1)$ symmetry; via the familiar Higgs mechanism, the Higgs field and vector field acquire the usual masses
\begin{equation} m_H = \mu\sqrt{2} \quad \text{and} \quad m_V = \frac{e\mu\sqrt{2}}{\sqrt{\lambda}}. \end{equation}
Ultimately, the Higgs mass sets the energy scale of the theory, making only $m_V/m_H$ physical (also making only the ratio $e/\sqrt{\lambda}$ relevant). 

The concept of a vortex is essentially a (2+1)-dimensional one, so we extend the concept to four spacetime dimensions by demanding our solutions be independent of $z$. Since a topological soliton is indexed by a mapping from spatial infinity to the internal vacua, we characterize the vortex by the homotopy group
\begin{equation} \pi_1\left(S^1\right) \cong \mathds{Z}. \end{equation}
A mapping that winds around the internal vacuum space (the circle here) $k$ times is said to correspond to a charge $k$ vortex. The greater charge vortices can be studied by constructing the moduli space based upon the charge $1$ solution. As in the main text, we take as our Ansatz for the static background
\begin{equation} \phi_{cl} = \frac{\mu}{\sqrt{\lambda}}e^{i\varphi}\Phi_{cl}(r) \quad \text{and} \quad A_{cl\mu} = -\frac{1}{e}\partial_{\mu}\varphi \ \mathscr{A}_{cl}(r), \end{equation}
where the asymptotic behavior of the one-form components at infinity is required to make the action finite. Based on this principle and smoothness at the origin, the profiles thus defined satisfy the conditions
\begin{equation} \lim_{|\mathbf{x}|\rightarrow 0} \Phi_{cl}(r) = \lim_{|\mathbf{x}|\rightarrow 0} \mathscr{A}_{cl}(r) = 0 \quad \text{and} \quad \lim_{|\mathbf{x}|\rightarrow \infty}\Phi_{cl}(r) = \lim_{|\mathbf{x}|\rightarrow \infty}\mathscr{A}_{cl}(r) = 1, \end{equation}
where $\mathbf{x}$ is in the transverse plane. The natural setting to describe this vortex is in cylindrical coordinates since we have $z$-translational invariance and the only nonvanishing one-form component is $A_{cl\varphi}$. The classical equations of motion in terms of these form factors are quite familiar ---
\begin{align}\label{eq1} -\frac{d}{dr}\left(\frac{1}{r}\frac{d\mathscr{A}_{cl}}{r}\right) - \frac{2e^2 \mu^2}{\lambda r}\Phi_{cl}^2(1-\mathscr{A}_{cl}) & = 0 \\ \label{eq2} -\frac{d}{dr}\left(r\frac{d\Phi_{cl}}{dr}\right) + \mu^2 r \Phi_{cl} (\Phi_{cl}^2 -1) + \frac{1}{r}\Phi_{cl}(1-\mathscr{A}_{cl})^2 & = 0. \end{align}

No analytic solution is known, but the form factors are shown numerically in Figure \ref{background} for $m_H = m_V$. The form factors are characterized by a width which goes like the inverse mass; effectively this vortex is nontrivial only in a compact core about the origin and extending infinitely in the $\pm z$-directions; it is in this sense we speak of a cosmic string. Since the fields for the static background are nondynamical, the Hamiltonian is particularly simple to compute; in terms of the form factors, the tension of the string is
\begin{equation}\label{tension} T_{\text{bkg}} = \frac{\pi}{e^2}\int dr \left[\frac{1}{r}\left(\frac{d\mathscr{A}_{cl}}{dr}\right)^2 + \frac{2re^2\mu^2}{\lambda}\left(\frac{d\Phi_{cl}}{dr}\right)^2 + \frac{2e^2\mu^2}{\lambda r}\Phi_{cl}^2(1-\mathscr{A}_{cl})^2\right]. \end{equation}

As one might expect, the specific case $m_H = m_V$ graphed in Figure \ref{background} holds special significance. The Hamiltonian written in \eqref{tension} can be cast in a different but illuminating way, utilizing the Bogomoln'yi trick \cite{Bogomoln'yi}. Keeping in mind that the metric, $g_{\mu \nu}$, is now in cylindrical coordinates, we may write the tension as 
\begin{multline}\label{BPS tension} T_{\text{bkg}} = \int d^2 x \sqrt{-g}\left[\frac{1}{4}g^{ik}g^{jl}F_{ij}F_{kl} - g^{ij}\nabla_i \phi_{cl}^* \nabla_j \phi_{cl} + \frac{\lambda}{2}\left(\phi^* \phi - \frac{\mu^2}{\lambda}\right)^2\right] \\ = \int d^2 x \sqrt{-g}\left[\frac{1}{2}\left(\frac{1}{r}\frac{dA_{cl\varphi}}{dr}\mp \sqrt{\lambda}\left(\phi^* \phi - \frac{\mu^2}{\lambda}\right)\right)^2 + \left|\nabla_r \phi_{cl} \mp \frac{i}{r}\nabla_{\varphi}\phi_{cl}\right|^2 + \mathscr{F}\left(\frac{m_V}{m_H}\right)\right], \end{multline}
where $\mathscr{F}$ is a function of the mass ratios and the fields. What is important is that the integral of $\mathscr{F}$ is a positive topological property of the string (that is, independent of the actual solution for $\Phi_{cl}$ and $\mathscr{A}_{cl\mu}$), which turns out to be a constant times the magnetic flux threaded through the core of the string (times the modulus of the charge number for a charge $k$ string), \emph{only} when $m_H = m_V$; see, for instance, \cite{Tong}. Thus the tension, and consequently the action, is minimized to this topological quantity only when the squared terms in \eqref{BPS tension} identically vanish. Hence we arrive at the BPS equations for the theory ---
\begin{align}\label{BPS 1} \left(\frac{d\mathscr{A}_{cl}}{dr}\right)^2 - \mu^4 r^2(\Phi_{cl}^2-1)^2 & = 0 \\ \label{BPS 2} r^2\left(\frac{d\Phi_{cl}}{dr}\right)^2 - \Phi_{cl}^2(1-\mathscr{A}_{cl})^2 & = 0. \end{align}
In this limit, we obtain first-order differential equations, which is a marked improvement. The solutions shown in Figure \ref{background} were constructed from the general equations of motion \eqref{eq1} and \eqref{eq2}. We then showed numerically that the resulting functions also satisfied these first order equations for $m_H = m_V$ as an after-the-fact test that these indeed are the correct BPS equations.

In the main text, we explain how the collective coordinates $X^i$ describing the embedding of the long string yield an effective theory whose leading action term is the Nambu-Goto action. In the 1960s, Callan, Coleman, Wess and Zumino (CCWZ) published a paper for constructing actions in the case of spontaneously broken internal symmetries based on a coset construction \cite{CCWZ}, which was quickly generalized to the case of broken spacetime symmetries, which we have here \cite{spacetime1, spacetime2}. Consider an infinitely long vortex appearing in a $d$ dimensional Lorentz invariant theory, which corresponds to the symmetry breaking $ISO(d-1,1)\longrightarrow ISO(1,1)\times O(d-2)$. The vacuum manifold corresponds to the coset constructed from these two spaces. The CCWZ construction introduces fields to parametrize this coset space, 
\begin{equation} \Omega = e^{i(\sigma^{\alpha}P_{\alpha}+X^i P_i)}e^{i\xi^{\alpha i}J_{\alpha i}}, \quad \Omega \in ISO(d-1,1)/ISO(1,1)\times O(d-2), \end{equation}
where $\alpha$ runs over the two vortex directions and $i$ runs over the $d-2$ transverse directions so that $P_i$ and $J_{\alpha i}$ are the broken generators. Then, one can decompose the Maurer-Cartan one-form constructed from such an $\Omega$ in terms of the available generators, i.e.
\begin{equation} \Omega^{-1} \partial_{\mu} \Omega = ie_{\mu}^{\phantom{\mu}\alpha}\left[P_{\alpha} + \nabla_{\alpha} X^i P_i + \nabla_{\alpha} \xi^{\beta i}J_{\beta i}\right] + i\omega_{\mu}^{\phantom{\mu}\alpha \beta}J_{\alpha \beta} + i\omega_{\mu}^{\phantom{\mu}ij}J_{ij}, \end{equation}
where $e_{\mu}^{\phantom{\mu}\alpha}$ is the coset vielbein, $\nabla_{\alpha} X^i$ and $\nabla_{\alpha} \xi^{\beta i}$ are derivatives covariant under $ISO(1,1)$ and $\omega_{\mu}^{\phantom{\mu}\alpha \beta}$ and $\omega_{\mu}^{\phantom{\mu}ij}$ are spin connections. The most general Lorentz invariant action describing the low-energy theory is then built from the above five geometric objects in combinations invariant under $ISO(1,1)$\footnote{The only subtlety is that $\xi^{\alpha i}$ can actually be expressed in terms of the $X^i$ via the so-called \emph{inverse Higgs constraint}. This is simply because the broken boosts/rotations can be re-expressed in terms of \emph{local} translations.}. The details can be found in many papers; a brief account is given in \cite{Low and Manohar} with details in their references. The generalization to supersymmetry and superstrings is given in \cite{Ivanov and Krivonos} and \cite{Hughes and Polchinski}. Doing this calculation for the bosonic case easily shows that the most relevant IR action term one can write down for this vortex theory is
\begin{equation} S = -\frac{1}{2\pi\alpha'}\int d^2 \sigma \sqrt{\det(\partial_{\alpha}X^{\mu}\partial_{\beta}X_{\mu})}, \end{equation}
where $X^{\mu} = \{\sigma^{\alpha}, X^i\}$. This is the classic derivation of effective string theory; performing the above Maurer-Cartan one-form decomposition to higher order then provides subleading corrections to the Nambu-Goto action, as mentioned in the text.

Finally, we present some of the longer expressions for the fluctuations about the static cosmic string background. The quadratic fluctuation operator mentioned in the text is defined by
\begin{equation} \mathscr{L} \supset \begin{pmatrix} \tilde{\xi}^* & \tilde{\xi} & \tilde{\chi}^{\mu} \end{pmatrix} \tilde{\Omega} \begin{pmatrix} \tilde{\xi} \\ \tilde{\xi}^* \\ \tilde{\chi}^{\nu} \end{pmatrix}. \end{equation}
A straightforward calculation shows that this quadratic operator is
\begin{equation}\label{fluctuation operator} \tilde{\Omega} = \begin{pmatrix} \tilde{\Omega}_1 & \tilde{\Omega}_2 & \tilde{\Omega}_3 \end{pmatrix}, \end{equation}
where
\begin{align} \tilde{\Omega}_1 & = \begin{pmatrix} \displaystyle -\frac{\square}{2}-ieA^{\mu}_{cl}\partial_{\mu} + \frac{e^2 A_{cl}^2}{2}-\lambda|\phi_{cl}|^2 + \frac{\mu^2}{2} \\ \displaystyle -\frac{\lambda}{2}(\phi_{cl}^*)^2 \\ \displaystyle e^2 \phi_{cl}^* A_{cl\mu} -ie\phi_{cl}^*\partial_{\mu} \end{pmatrix}, \\ \tilde{\Omega}_2 & = \begin{pmatrix} \displaystyle -\frac{\lambda}{2}(\phi_{cl})^2 \\ \displaystyle -\frac{\square}{2}+ieA_{cl}^{\mu}\partial_{\mu}+\frac{e^2 A_{cl}^2}{2}-\lambda|\phi_{cl}|^2+\frac{\mu^2}{2} \\ \displaystyle e^2 \phi_{cl} A_{cl\mu} +ie\phi_{cl}\partial_{\mu} \end{pmatrix}, \\ \text{and} \quad \tilde{\Omega}_3 &= \begin{pmatrix} \displaystyle e^2 \phi_{cl} A_{cl\nu} -ie\partial_{\nu}\phi_{cl} \\ \displaystyle e^2 \phi_{cl}^* A_{cl\nu}+ie\partial_{\nu}\phi_{cl}^* \\ \displaystyle \frac{\eta_{\mu \nu}}{2}\square-\frac{\partial_{\mu}\partial_{\nu}}{2}+e^2\eta_{\mu \nu}|\phi_{cl}|^2 \end{pmatrix}. \end{align}
This operator is indeed Hermitian in the sense of integration by parts in the same way that the momentum $P_{\mu} = i\partial_{\mu}$ is. The fluctuation equations of motion derived from this operator are then
\begin{align}\label{eom 1} ier\phi_{cl}\bar{\xi}_{m-2}-ier\phi_{cl}^* \xi_m + \frac{i(m-1)}{r}\chi_{(m-1)\varphi} + \frac{d}{dr}\left(r\chi_{(m-1)r}\right) & = 0  \\ \notag \left[-ier\partial_r \phi_{cl}^* + ier\phi_{cl}^* \frac{d}{dr}\right]\xi_m + \left[ier\partial_r \phi_{cl} - ier\phi_{cl} \frac{d}{dr}\right]\bar{\xi}_{m-2}-\left[\frac{i(m-1)}{r}\frac{d}{dr}\right]\chi_{(m-1)\varphi} & \\ \label{eom 2} - \left[\frac{(m-1)^2}{r}+2e^2 r \phi_{cl}^* \phi_{cl} - M^2 r\right]\chi_{(m-1)r} & = 0  \\ \notag  \left[-e\phi_{cl}^* - 2e^2 \phi_{cl}^* A_{cl\varphi} - em\phi_{cl}^*\right]\xi_m + \left[-e\phi_{cl}-2e^2 \phi_{cl}A_{cl\varphi}+e(m-2)\phi_{cl}\right]\bar{\xi}_{m-2}    & \\ \label{eom 3} +\left[M^2-2e^2 \phi_{cl}^*\phi_{cl}\right]\chi_{(m-1)\varphi}+r\frac{d}{dr}\left(\frac{1}{r}\frac{d\chi_{(m-1)\varphi}}{dr}\right) -i(m-1)\frac{d}{dr}\left(\frac{1}{r}\chi_{(m-1)r}\right) & = 0   \\ \label{eom 4} \frac{1}{r}\frac{d}{dr}\left(r\frac{d\chi_{(m-1)3}}{dr}\right)+\left(M^2 -\frac{(m-1)^2}{r^2}-2e^2\phi_{cl}^*\phi_{cl}\right)\chi_{(m-1)3}  & = 0,\end{align}
where the first and last equations are strictly valid only for $M\neq 0$. By abuse of notation, the $\phi_{cl}^*$ and $\phi_{cl}$ factors above do \emph{not} carry the phases; they are distinguished only to show the parallel between the different terms.


\acknowledgments

I am very grateful to Raphael Flauger who guided me and oversaw this work, without whom this paper would not be possible. I would also like to thank Ira Rothstein for providing stimulating discussions on some of the topics.




\end{document}